\documentclass[a4paper,11pt]{article}
\pdfoutput=1 
\usepackage{graphicx}
\usepackage{amsmath}
\usepackage{bm}
\usepackage{xcolor}
\usepackage{jcappub} 
\usepackage{subfig}

\usepackage[T1]{fontenc} 
\def\vq{\vec{q}}
\def\vk{\vec{k}}
\def\ft{\tilde{f}}
\def\gt{\tilde{g}}
\def\Zb{\bar{Z}}
\def\Yb{\bar{Y}}
\def\Xb{\bar{X}}
\def\Wb{\bar{W}}

\def\Sch{\text{Schr\"{o}dinger}\ }
\def\Mpl{M_{\rm Pl}}

\title{\boldmath  Non-Gaussianity from Entanglement During Inflation }
\author[a,d]{Nadia Bolis,}
\author[a]{Andreas Albrecht,}
\affiliation[a]{Center for Quantum Mathematics and Physics, and Department of Physics, University of California at Davis, One Shields Ave, Davis, CA 95616, USA}

\author[b]{R.~Holman}

\affiliation[b]{Division of Business and Computational Studies, Minerva Schools at KGI, 1145 Market Street, San Francisco, CA 94103, USA}
\affiliation[c]{%
Central European Institute for Cosmology and Fundamental Physics, Institute of
Physics, Czech Academy of Sciences, Na Slovance 1999/2 Prague, Czech Republic}
\emailAdd{nbolis@ucdavis.edu}
\emailAdd{ajalbrecht@ucdavis.edu}
\emailAdd{rholman@minerva.kgi.edu}

\abstract{We compute the bi-spectrum of CMB temperature fluctuations for a state where the metric perturbation $\zeta$ is entangled with a spectator scalar field $\chi$. Novel terms in the cubic $\zeta$ action coupled to the scalar can be the dominant contribution to the bi-spectrum for such states and we highlight the differences between this result and the no-entanglement bi-spectrum. New shapes can be important in the bi-spectra leading to distinctive observational signatures.}

\begin{document}
\maketitle
\flushbottom

\section{Introduction\label{sec:intro}}

A sufficient amount of cosmic inflation will expand microscopic features of the state of the universe to all observable scales.  The canonical assumption, that on small scales the initial state of inflation is described by the Bunch-Davies (BD) vacuum, has produced extremely successful predictions~\cite{Ade:2015xua}.  
In ~\cite{Albrecht:2018hoh} we have advocated the use of cosmic inflation as a probe of the properties of the initial state ("the most powerful microscope in the Universe"), and showcased specific possible deviations for the BD vacuum due to quantum entanglement which could be tested in this manner. 

Entangled states involve allowing other fields, such as spectator scalars\cite{Albrecht:2014aga} or tensor metric perturbations\cite{Bolis:2016vas} to become entangled with the scalar fluctuations in the quantum state. A generic effect of such entanglement is the inducing of oscillations in the various power spectra of the CMB which are already well constrained by current data\cite{Ade:2015lrj,Akrami:2018odb}. A possible origin of such entangled states is discussed in \cite{Bolis:2018jmo,Holman:2019spa}.

In prior work~\cite{Albrecht:2014aga,Bolis:2016vas} we calculated the predictions at the level of power spectra for various examples of initial state entanglement and found interesting signatures which may already be present in subtle features of the data (a more systematic analysis is underway~\cite{ABA}).

The power spectrum is mainly a probe of the linearized sector of field fluctuations. It can also probe interactions at loop level, although these effects will typically be suppressed. To understand the interacting aspects of the theory and to have a chance to lift the degeneracies between different inflationary models, we need to compute higher point functions, such as the bi and tri spectra. Data on these correlation functions has been gathered from both the CMB and large scale structure (LSS). While LSS non-Gaussianity data has great potential for improvements \cite{Alvarez:2014vva}, CMB data, in particular from the Planck collaboration \cite{Ade:2015ava}, provides us the best bounds to date. 

%To search for a non-Gaussian signal, theoretically motivated shapes of bispectra and trispectra are compared to observational templates. The most common bispectra shapes looked for are the equilateral, orthogonal and local triangle configurations. Higher derivative single field models, such as `k-inflation' [cite], Galileon-like models[cite], Horndeski models [cite] and ghost-inflation[cite], produce a higher equilateral or orthogonal signal. Local non-Gaussianity, on the other hand, can arise from multi-field inflationary models.  The extra degrees of freedom of the other field(s) in multi-field will generally produce isocurvature modes, which can generate non-Gaussianities in the curvature perturbation $\zeta$. 

In this paper we determine the shape of the bi-spectrum produced by entanglement during inflation and compare it to the no-entanglement BD bi-spectrum. We also find that the shape produced by entanglement differs from both the shape produced by an alternative non-BD model \cite{Holman:2007na} (for non-Gaussianity produced by other models with an excited state or extra degrees of freedom see \cite{Agullo:2010ws,
Flauger:2016idt,
Meerburg:2009ys,
Kundu:2013gha}) and that of many multi-field inflation models (i.e local shape)\cite{Wands:2010af}. Part of what drives this difference is the fact that there are new cubic operators contributing to the bi-spectrum that would not have contributed for the BD state. These operators are of the schematic form $\chi^2 \zeta$ with various operators appearing. These contribute because the state now depends on the spectator scalar field through entanglement, and the contribution is proportional to the number of scalar fields. Thus, these operators can give rise to new shapes {\em and} can be made to dominate over the operators giving the standard contribution to the bi-spectrum. 
We also find that the magnitude of the bi-spectrum for entangled states depends on how long inflation lasted. Non-BD states typically suffer from a back-reaction problem due to the energy density of what can be thought of as particles built upon the BD vacuum. The longer the inflationary period, the larger this effect and hence there is a strong motivation to stay near the minimal number of e-folds needed to solve the horizon and flatness problems. The same principle holds for this model of entanglement and therefore, this may provide natural limits to how strong the enhancements can be.

%in two ways. Either the super-Hubble isocurvature perturbations get converted to adiabatic curvature perturbations through a non-linear transfer mechanism, or there can be non-Gaussianities inherent in the isocurvature modes [cite]. 

%\textcolor{red}{maybe this paragraph is out of place here} Local type non-Gaussianity is maximized in the squeezed limit ($k_3<<k_1\approx k_2$), where the largest wavelength mode acts as a slowly varying background to the shorter wavelength modes. As shown in \cite{Creminelli:2004yq} an invaluable consistency relation follows: single field slow roll inflation cannot produce a large squeezed non-Gaussian signal. With the only assumptions that inflation is driven by a single field and starts in the BD vacuum, regardless of whether it is a higher derivative theory or the exact form of the potential, the squeezed bi-spectrum will be proportional to $1-n_s$, where $n_s$ is the spectral index. Therefore, for a nearly scale invariant power spectrum, the squeezed bi-spectrum will be highly suppressed. This result is especially important because it means that a large signal of local non-Gaussianity, would effectively rule out all these single field models, including the simplest models preferred today. As it stands, however, the current bounds on local non-Gaussianity are still consistent with single field models. From the Planck Collaboration $f_{NL}^{\text{loc}} = 2.5 \pm 5.7$ for temperature data and $f_{NL}^{\text{loc}} = 0.8 \pm 5.0$ for temperature and polarization data \cite{Ade:2015ava}. 
 
In section \ref{sec:2} we first give a brief review of the construction of entangled states using the field theoretic Schr\"{o}dinger picture in the Gaussian approximation. We then extend this method to allow for cubic deviations from Gaussianity. We use cosmological perturbation theory, in the presence of the relevant cubic Hamiltonians, to compute the bi-spectrum in terms of  the cubic coefficient functions of the  Schr\"{o}dinger wavefunctional.

%Then we introduce the  approach to calculating the bi-spectrum and use this formalism to find the general bi-spectrum for this entangled state as well as the equations of motion of the cubic order state coefficients. 

In section \ref{sec:3} we follow this use of  cosmological perturbation with a second perturbative expansion in the entanglement strength parameter. This parameter is necessarily constrained to be small from observations of the CMB power spectrum so that a perturbative approach is valid here.

%We find that terms previously neglected in the cubic Hamiltonian coupling $\zeta$ to the spectator scalar $\chi$ take on a preeminent role in our calculation due to the fact that their effects can be enhanced by having many scalars appear. Interestingly such enhancements do {\em not} appear in the calculation of the entangled CMB power spectrum. 

In section \ref{sec:4} we present the resulting bi-spectrum generated by these extra terms and the entanglement. We plot the shape function of this bi-spectrum and look at its equilateral, flattened and squeezed triangle limits.  In section \ref{sec:5} we present our conclusions. 

\section{\label{sec:2}Entanglement in the Schr\"{o}dinger Picture}

The entangled states we use in this work\cite{Albrecht:2014aga} are best understood within the Schr\"{o}dinger picture field theory formalism. Here we think of a wave-functional $\Psi\left[\zeta(\cdot), \chi(\cdot); \tau\right]$ which gives the probability amplitude for finding the scalar metric perturbation $\zeta$ and spectator $\chi$ field configurations on the spatial hypersurface specified by the conformal time $\tau$. Note that in this formalism, $\zeta(\cdot), \chi(\cdot)$ are time independent; $(\cdot)$ denotes the slot for the spatial variables. To compute observables we first solve the \Sch equation for $\Psi\left[\zeta(\cdot), \chi(\cdot); \tau\right]$ and then use this wave-functional to compute expectation values.

\subsection{\label{subsec:quad}The Gaussian Approximation}

We recapitulate the discussion in ref~\cite{Albrecht:2014aga} using the continuum momentum space rather than the box normalized one used previously.

Much as any other non-trivial quantum mechanics problem, we start from the exactly soluble Gaussian problem of the free theory and then perturb around it. The quadratic action for the $\zeta$-$\chi$ system is:

\begin{equation}
\label{eq:zetachiquad}
S_2 = \int d^4x\ a(\tau)^2\left\{z\left(\zeta^{\prime 2}-c_s^2 (\partial_i\zeta)^2\right)+\frac{1}{2}\left(\chi^{\prime 2}-(\partial_i \chi)^2-a(\tau)^2 m_{\chi}^2 \chi^2\right)\right\},
\end{equation}
where $a(\tau)$ is the scale factor, $z=\epsilon \Mpl^2\slash c_s^2$, $\epsilon$ is the slow-roll parameter and $c_s\leq 1$ is the sound speed of metric perturbations. From this we find the canonically conjugate momenta and the quadratic Hamiltonian:

\begin{equation}
\label{eq:canonicalmom}
\Pi_{\zeta} = 2 z a(\tau)^2 \zeta^{\prime}, \quad \Pi_{\chi} = a(\tau)^2 \chi^{\prime},
\end{equation}

\begin{eqnarray}
\label{eq:quadHam}
& & H_2= \int d^3 x\ \left\{\frac{\Pi_{\zeta}^2}{4 z a(\tau)^2}+\frac{\Pi_{\chi}^2}{2 a(\tau)^2}+(z c_s^2 a(\tau)^2) (\partial_i \zeta)^2+\frac{1}{2} a(\tau)^2 (\partial_i \chi)^2+\frac{1}{2} a(\tau)^4 m_{\chi}^2 \chi^2\right\}\nonumber \\
&=& \int \frac{d^3 k}{(2\pi)^3}\ \left\{\frac{\Pi_{\zeta \vec{k}}\Pi_{\zeta -\vec{k}}}{4 z a(\tau)^2}+\frac{\Pi_{\chi \vec{k}}\Pi_{\chi -\vec{k}}}{2 a(\tau)^2}+(z c_s^2 k^2 a(\tau)^2)\zeta_{\vec{k}} \zeta_{-\vec{k}}
+\frac{1}{2}  a(\tau)^2 (k^2+a^2(\tau) m_{\chi}^2) \chi_{\vec{k}}\chi_{-\vec{k}}\right\},\nonumber\\
\end{eqnarray}
where we've decomposed the fields into spatial momentum modes in the second equation:

\begin{equation}
\label{eq:momdecomp}
\zeta(\vec{x}) = \int \frac{d^3 k}{(2\pi)^3}\ \zeta_{\vec{k}}e^{-i \vec{k}\cdot\vec{x}},\ \Pi_{\zeta}(\vec{x})=\int \frac{d^3 k}{(2\pi)^3}\ \Pi_{\zeta \vec{k}}e^{-i \vec{k}\cdot\vec{x}},
\end{equation}
and likewise for $\chi,\ \Pi_{\chi}$.

The Gaussian ansatz we use that allows for entanglement between $\zeta$ and $\chi$ is:

\begin{equation}
\label{eq:gaussentansatz}
\Psi_G\left[\zeta,\chi; \tau\right] = N(\tau)\exp\left[-\frac{1}{2} \Big(\langle A(\tau) \zeta \zeta\rangle+\langle B(\tau) \chi \chi\rangle+\langle C(\tau) \left(\zeta \chi+\chi \zeta\right)\rangle\Big)\right],
\end{equation}
where our notation in general is:

\begin{equation}
\label{eq:momintnotation}
\langle S \phi_1\ldots \phi_n\rangle \equiv \int \left(\prod_{i=1}^n \frac{d^3 k_i}{(2\pi)^3}\right)\ (2\pi)^3 \delta^{(3)}\left(\sum_{j=1}^n \vec{k}_i\right)\ S(\vec{k}_1,\ldots,\vec{k}_n) \phi_1(\vec{k}_1)\ldots \phi_n(\vec{k}_n).
\end{equation}

Inserting this ansatz into the \Sch equation $i\partial_{\tau}\Psi_G = H_2 \Psi_G$, using  the following commutation relations:
\begin{equation}
\label{eq:commrels}
\left[\Pi_{\zeta \vec{k}}, \zeta_{\vec{q}}\right]=-i (2\pi)^3 \delta^{(3)}(\vec{k}+\vec{q}),\quad \left[\Pi_{\chi \vec{k}}, \chi_{\vec{q}}\right]=-i (2\pi)^3 \delta^{(3)}(\vec{k}+\vec{q})
\end{equation}
to identify $\Pi_{\zeta \vec{k}}=-i(2\pi)^3\delta\slash \delta \zeta_{-\vec{k}}$ (and likewise for $\chi$) and matching powers of the modes yields\cite{Albrecht:2014aga}:
\begin{eqnarray}
\label{eq:quadkernels}
 i A_k'(\tau)&=&\left(\frac{A_k^2}{2 z a^2}+\frac{C_k^2}{a^2}\right)-2 z c_s^2 a^2 k^2\nonumber\\
 i B_k'(\tau)&=&\left(\frac{B_k^2}{a^2}+\frac{C_k^2}{2 z a^2}\right)- a^2 (k^2+a^2 m_{\chi}^2)\nonumber\\
 i\frac{C_k'(\tau)}{C_k(\tau)} &=&\left(\frac{A_k}{2 z a^2}+\frac{B_k}{a^2}\right)\nonumber\\
 i\frac{N'(\tau)}{N(\tau)} &=&\left[(2\pi)^3\delta^{(3)}(\vec{q}=\vec{0})\right]\int \frac{d^3 k}{(2\pi)^3}\left(\frac{A_k}{2 z a^2}+\frac{B_k}{a^2}\right).
\end{eqnarray}
We have defined $A_k\equiv A(\vec{k},-\vec{k})$ and likewise for the other kernels. We have also made use of the fact that rotational invariance dictates that these kernels will only depend on the magnitude $k$ of $\vec{k}$. Note the appearance of $(2\pi)^3\delta^{(3)}(\vec{q}=\vec{0})$ in the equation for the normalization factor; this is the comoving volume ${\cal V}$ of the spatial box we quantize the system in. We can neglect this by dividing by the normalization factor $\langle \Psi_G|\Psi_G\rangle$ when calculating observables. 

The equations for the kernels $A_k,\ B_k$ are Riccatti type equations and can be converted from a set of non-linear first equations to linear second order ones. Setting 

\begin{eqnarray}
\label{eq:Riccatti}
i A_k(\tau) &=& \alpha(\tau)^2 \left(\frac{f_k'(\tau)}{f_k(\tau)}-\frac{\alpha'(\tau)}{\alpha(\tau)}\right)\nonumber\\
i B_k(\tau) &=& a(\tau)^2 \left(\frac{g_k'(\tau)}{g_k(\tau)}-\frac{a'(\tau)}{a(\tau)}\right),
\end{eqnarray}
with the definition of $\alpha^2 = 2 z a^2$. Doing this converts eqs.\eqref{eq:quadkernels} into

\begin{eqnarray}
\label{eq:riccatticonvert}
&& f_k''(\tau)+\left(c_s^2 k^2 -\frac{\alpha''}{\alpha}\right)f_k(\tau)=\frac{C_k^2}{\alpha^2 a^2}f_k(\tau)\nonumber\\
&& g_k''(\tau)+\left(\omega_{\chi k}^2 -\frac{a''}{a}\right)g_k(\tau)=\frac{C_k^2}{\alpha^2 a^2}g_k(\tau)\nonumber\\
&& \frac{C_k'(\tau)}{C_k(\tau)}=-\left( \frac{f_k'(\tau)}{f_k(\tau)}+\frac{g_k'(\tau)}{g_k(\tau)}-\frac{\alpha'(\tau)}{\alpha(\tau)}-\frac{a'(\tau)}{a(\tau)}\right),
\end{eqnarray}
where $\omega_{\chi k}^2\equiv k^2+a^2 m_{\chi}^2$.

We can solve the equation for the entanglement kernel $C_k$: $C_k(\tau) = \lambda_k(\alpha(\tau) a(\tau))\slash (f_k(\tau) g_k(\tau))$. The parameter $\lambda_k$ measures the entanglement strength and as shown in ref~\cite{Albrecht:2014aga}, requiring $\Psi_G$ to be normalizable, implies  $|\lambda_k|<1\slash 2$. Consistency with CMB data would likely also require $\lambda_k\lesssim 0.1$, and we will make use of this to simplify the calculation of the bi-spectrum by expanding the modes in a series of $\lambda_k$. We see from eqs.\eqref{eq:riccatticonvert} that the leading correction to the modes $f_k,\ g_k$ due to entanglement is of order $\lambda_k^2$. The same is true of the two point functions $\langle \zeta_{\vec{k}} \zeta_{-\vec{k}}\rangle,\langle \chi_{\vec{k}} \chi_{-\vec{k}}\rangle$ calculated from $\Psi_G$:

\begin{eqnarray}
\label{eq:twopointfuncs}
\langle \zeta_{\vec{k}} \zeta_{-\vec{k}}\rangle &=&\frac{B_{kR}}{A_{kR}B_{kR}-C_{kR}^2}\nonumber\\\langle \chi_{\vec{k}} \chi_{-\vec{k}}\rangle &=&\frac{A_{kR}}{A_{kR}B_{kR}-C_{kR}^2}\nonumber\\
\left \langle \left(\zeta_{\vec{k}} \chi_{-\vec{k}}+\zeta_{-\vec{k}} \chi_{\vec{k}}\right)\right \rangle &=&\frac{2 C_{kR}}{A_{kR}B_{kR}-C_{kR}^2}.\nonumber\\
\end{eqnarray}
Note that the cross field correlator is linear in $\lambda_k$ and vanishes when there is no entanglement. 

\subsection{\label{subsec:beyondquad}Beyond the Gaussian Approximimation}

To compute the three point function in the \Sch theory, we need to go beyond the Gaussian approximation of the previous subsection. The appropriate tool for this task is \Sch perturbation theory. 

We take the Hamiltonian to be of the form: $H= H_2+\mu H_3$, where $H_3$ consists of terms in the full $\zeta$-$\chi$ action that are of cubic order in the fields. The $\zeta$ contribution to $H_3$ is that calculated originally in ref.\cite{Maldacena:2002vr} and generalized to Horndeski theories in ref.\cite{Horndeski:1974wa}. There are also terms consisting of one power of $\zeta$ and two of $\chi$ as in refs.\citep{Weinberg:2005vy,delRio:2018vrj}. The parameter $\mu$ serves as the expansion parameter controlling the perturbative approximation. It can be made more explicit via arguments such as those in ref.\cite{Adshead:2017srh}, but we leave its exact specification open for now.

Following the ideas in ref.\cite{Collins:2017haz}, we generalize the Gaussian wave-functional to allow for a non-trivial three-point function:
\begin{eqnarray}
\label{eq:cubicwf}
\Psi\left[\zeta, \chi; \tau\right]&=&(1+\mu \Delta\left[\zeta,\chi; \tau\right])\Psi_G\left[\zeta, \chi; \tau\right]\nonumber\\
\Delta\left[\zeta, \chi; \tau\right]&=& \langle Z(\tau) \zeta \zeta \zeta\rangle+\langle Y(\tau) \chi \chi \chi\rangle+\sum_i \left(\langle W_i(\tau) \chi_i \zeta  \zeta\rangle +\langle X_i(\tau)\zeta_i \chi \chi  \rangle\right),
\end{eqnarray}
where we again use the abbreviated momentum integral notation of eq.\eqref{eq:momintnotation}. The sum over $i$ in $\Delta$ accounts for each of the three momenta matched with a given mode:
\begin{eqnarray}
\sum_i \langle W_i(\tau) \chi_i \zeta  \zeta\rangle &\equiv&  \int \left(\prod_{j=1}^n \frac{d^3 k_j}{(2\pi)^3}\right)\ (2\pi)^3 \delta^{(3)}\left(\sum_{j=1}^n \vec{k}_i\right)\ \left(W_1(\vk_1,\vk_2,\vk_3; \tau) \chi_{\vk_1}\zeta_{\vk_2}\zeta_{\vk_2}\right .\nonumber\\
&&\left .+W_2(\vk_1,\vk_2,\vk_3; \tau) \zeta_{\vk_1}\chi_{\vk_2}\zeta_{\vk_3}+W_3(\vk_1,\vk_2,\vk_3; \tau) \zeta_{\vk_1}\zeta_{\vk_2} \chi_{\vk_3}\right).
\end{eqnarray}
The $W_i$'s are symmetric in the momenta corresponding to the two $\zeta$ modes it multiplies and likewise for $X_i$ and the $\chi$ modes. The $Z,\ Y$ kernels are fully symmetric in their momentum arguments.

We now fix the various kernels in $\Delta$ using perturbation theory with $H_3$ acting as a perturbation on top of the quadratic Hamiltonian $H_2$. The parameter $\mu$ serves as the expansion parameter and we solve the \Sch equation order by order in $\mu$:
\begin{equation}
i\partial_{\tau} \Psi = H\Psi\Rightarrow i\partial_{\tau} \Psi_G = H_2 \Psi_G;\quad \left(i\partial_{\tau} \Delta\right)\Psi_G+\Delta i\partial_{\tau} \Psi_G = H_2 (\Delta \Psi_G)+H_3 \Psi_G.
\end{equation}
If we use the \Sch equation for $\Psi_G$ we can rewrite that for $\Delta$ as
\begin{equation}
\label{eq:delta}
\left(i\partial_{\tau} \Delta\right)\Psi_G=\left[H_2,\Delta\right] \Psi_G+H_3 \Psi_G.
\end{equation}
Using our expressions for $\Delta$ in eq.\eqref{eq:cubicwf} as well as that for the quadratic Hamiltonian, we find the following set of equations for the kernels:
\begin{eqnarray} \label{eq:cubickerneleoms}
i \partial_{\tau}Z^{\vk_1,\vk_2,\vk_3 } &=& Z^{\vk_1,\vk_2,\vk_3 } \left(\sum_{i=1}^3 \frac{A_{k_i}}{\alpha^2}\right) +\frac{1}{a^2}\sum_{i=1}^3 W^{\vk_1,\vk_2,\vk_3}_{i} C_{k_i}+\mathcal{S}^{\zeta_{\vk_1} \zeta_{\vk_2}\zeta_{\vk_3}}_Z\\
i \partial_{\tau}Y^{\vk_1,\vk_2,\vk_3} &=& Y^{\vk_1,\vk_2,\vk_3}\sum_{i=1}^3 \frac{B_{k_i}}{a^2} +\frac{1}{\alpha^2}\sum_{i=1}^3 X^{\vk_1,\vk_2,\vk_3}_{i} C_{k_i}+\mathcal{S}^{\chi_{\vk_1} \chi_{\vk_2} \chi_{\vk_3}}_Y\\
i\partial_{\tau}W_{i}^{\vk_1,\vk_2,\vk_3 } &=& W^{\vk_1,\vk_2,\vk_3}_{i} \left(\frac{B_{k_i}}{a^2}+\sum_{j\neq i}\frac{A_{k_j}}{\alpha^2}\right) +\frac{Z^{\vk_1,\vk_2,\vk_3}}{\alpha^2} C_{k_i} + \sum_{j\neq i, l\neq,i,j}\frac{X^{\vk_1,\vk_2,\vk_3}_{j}}{a^2} C_{k_l}+\nonumber\\
&+& \mathcal{S}^{\chi_{\vk_i} \zeta_{\vk_j}\zeta_{\vk_l}}_{W_i} \\
i\partial_{\tau}X_{i}^{\vk_1,\vk_2,\vk_3} &=& X^{\vk_1,\vk_2,\vk_3}_{i} \left(\frac{A_{k_i}}{\alpha^2}+\sum_{j\neq i}\frac{B_{k_j}}{a^2}\right) +\frac{Y^{\vk_1,\vk_2,\vk_3}}{a^2} C_{k_i} + \sum_{j\neq i, l\neq,i,j} \frac{W^{\vk_1,\vk_2,\vk_3}_{j}}{\alpha^2} C_{k_l}+\nonumber\\
&+& \mathcal{S}^{\zeta_{\vk_i} \chi_{\vk_j}\chi_{\vk_l}}_{X_i}
\end{eqnarray}
where $ \mathcal{S}^{\zeta_{\vk_1} \zeta_{\vk_2} \zeta_{\vk_3}}_Z, \mathcal{S}^{\chi_{\vk_1} \chi_{\vk_2} \chi_{\vk_3}}_Y, \mathcal{S}^{\chi_{\vk_i} \zeta_{\vk_j}\zeta_{\vk_l}}_{W_i}$ and $\mathcal{S}^{\zeta_{\vk_i} \chi_{\vk_j}\chi_{\vk_l}}_{X_i}$ are the source terms derived from $H_3 \Psi_G$  that correspond to each given combination of field modes.

There are two cubic Hamiltonians that contribute to the source term $H_3 \Psi_G$. First, the cubic Hamiltonian for $\zeta$, which can be calculated from the cubic order $\zeta$ action \cite{Agarwal:2013rva}:
\begin{equation}
S^{\zeta^3}_3 = \int dx^{(3)}dt \Big[ -\frac{2\lambda_c}{\Sigma} \frac{a^3 \epsilon}{c_s^2H} \dot{\zeta}^3 - \gt \dot{\zeta}(\partial_i \zeta)^2 +\frac{a^3\epsilon}{c_s^2}(2s+\epsilon -\eta) \zeta \dot{\zeta}^2 + a \epsilon (\epsilon +\eta) \zeta (\partial_i \zeta)^2- 2 \frac{a^3 \epsilon^2}{c_s^4} \dot{\zeta}\partial_i \zeta \partial^i \partial^{-2}\dot{\zeta}\Big].
\end{equation}
Using the conjugate momenta $\Pi^{\zeta}_k$ in eq.\eqref{eq:canonicalmom}, the cubic $\zeta$ Hamiltonian is:

\begin{eqnarray}
H_3^{\zeta^3} &=& - \int \prod_i^3 d^3\vec{k}_i\ \delta(\sum_j \vec{k}_j)\ \Big[ \frac{1}{a^2 \epsilon} \frac{\epsilon- \eta}{8}\left\{\zeta_1 \Pi^{\zeta}_2 \Pi^{\zeta}_3 +  \Pi^{\zeta}_2 \Pi^{\zeta}_3  \zeta_1 + {\rm permutations}. \right\} \\\nonumber
&-& a^2 \frac{\epsilon(\epsilon +\eta)}{3}\left\{\vec{k}_2 \cdot \vec{k}_3 + p.m. \right\}  \zeta_1 \zeta_2 \zeta_3 \Big] 
- \frac{1}{a^2} \frac{1}{12}  \left\{\frac{\vec{k}_2 \cdot \vec{k}_3}{k_3^2} \left[ \Pi^{\zeta}_1 \zeta_2 \Pi^{\zeta}_3 + \Pi^{\zeta}_3 \zeta_2 \Pi^{\zeta}_1\right] +{\rm permutations} \right\}
\end{eqnarray}
%\begin{eqnarray} \label{ham}
%H_{\zeta}^{(3)} &=& - \int \prod_i^3 d\vk_i \delta\big(\sum_j \vk_j \big) \Big[-\frac{2\lambda_c}{\Sigma}\frac{c_s^4}{a^5 \epsilon} \Pi_{\vk_1}\Pi_{\vk_2}\Pi_{\vk_3}+\gt \frac{c_s^2}{a^3\epsilon}\frac{1}{3} \Big\{\frac{\vk_2 \cdot \vk_3}{2} \Big[ \Pi_{\vk_1} \zeta_{\vk_2} \zeta_{\vk_3}+\zeta_{\vk_2} \zeta_{\vk_3}\Pi_{\vk_1} \Big] + p.m. \Big\} \nonumber\\
%&+& \frac{c_s^2}{a^2\epsilon} (2s+\epsilon -\eta) \Big\{\frac{1}{2}\Big[ \zeta_{\vk_1}\Pi_{\vk_2}\Pi_{\vk_3}+\Pi_{\vk_2}\Pi_{\vk_3}\zeta_{\vk_1}\Big]+p.m.\Big\} + a^2\epsilon(\epsilon+\eta) \frac{1}{3} \Big\{-\vk_2 \cdot \vk_3 \zeta_{\vk_1}\zeta_{\vk_2}\zeta_{\vk_3}+p.m.\Big\}\nonumber\\
%&\textcolor{red}{-}&\frac{2}{a^2} \frac{1}{3} \Big\{-\frac{\vk_2 \cdot \vk_3}{2k_3^2}\Big[\Pi_{\vk_1}\zeta_{\vk_2}\Pi_{\vk_3}+\Pi_{\vk_3} \zeta_{\vk_2}\Pi_{\vk_1} \Big]+p.m.\Big\} \Big]
%\end{eqnarray}
where we have symmetrized the mixed conjugate momentum and $\zeta$ terms. We also choose to take $c_s=1$ for simplicity so the first two terms from the action vanish. 

The second Hamiltonian does not contribute to the BD bi-spectrum but is relevant here. It involves interaction terms that are schematically of the form $\zeta \chi^2$\cite{Weinberg:2005vy,delRio:2018vrj}. Their explicit form is:
\begin{eqnarray}
H_3^{\zeta \chi^2} &=& - \int \prod_i^3 d^3\vec{k}_i\; \delta(\sum_j \vec{k}_j) \Big[ \frac{\epsilon}{2 a^2 } \frac{1}{2}\left\{\zeta_1 \Pi^{\chi}_2 \Pi^{\chi}_3 +  \Pi^{\chi}_2 \Pi^{\chi}_3  \zeta_1 + {\rm permutations} \right\} \\\nonumber
&-& \frac{\epsilon a^2}{2} \frac{1}{3}\left\{\vec{k}_2 \cdot \vec{k}_3 + {\rm permutations}. \right\}  \zeta_1 \chi_2 \chi_3 \Big] 
- \frac{1}{ a^2} \frac{1}{12}  \left\{\frac{\vec{k}_2 \cdot \vec{k}_3}{k_3^2} \left[ \Pi^{\chi}_1 \chi_2 \Pi^{\zeta}_3 + \Pi^{\zeta}_3 \chi_2 \Pi^{\chi}_1\right] +{\rm permutations} \right\}.
\end{eqnarray}

These will contribute to source terms for the cubic kernels, which in turn will contribute to the bi-spectrum. These terms can dominate over pure $\zeta^3$ terms due to the fact that they are proportional to the number of scalars present; because of this, and the fact that their contributions to the bi-spectrum are qualitatively different from the BD ones, we will focus our attention on these terms in the remainder of this paper.

\subsection{The Schr\"{o}dinger Picture Approach to Calculating the Bi-spectrum}

In this subsection we introduce the Schr\"{o}dinger picture approach to calculating the bi-spectrum. This is a novel use of Schr\"odinger field theory. It gives the same answer as the interaction picture calculation\cite{Maldacena:2002vr}, but generalizes to states such as the entangled one used here in a more intuitive way than is possible in the interaction picture.

The entangled three-point function is defined by taking the expectation value of $\langle\zeta_{\vk_1} \zeta_{\vk_2} \zeta_{\vk_3}\rangle$ using the  entangled cubic state of eq.\eqref{eq:cubicwf}:
\begin{equation}
\label{eq:3point}
\langle\zeta_{\vk_1} \zeta_{\vk_2} \zeta_{\vk_3}\rangle=\int \mathcal{D}^2 \zeta_{\vq}\  \mathcal{D}^2 \chi_{\vq}\ \left\{ (1+\mu \Delta^*\left[\zeta,\chi; \tau\right])\ \zeta_{\vk_1}\zeta_{\vk_2} \zeta_{\vk_3}\ (1+\mu \Delta\left[\zeta,\chi; \tau\right])\ |\Psi_G|^2\right\}
\end{equation}
where $\Delta\left[\zeta,\chi; \tau\right]$ is defined in eq.\eqref{eq:cubicwf} and the Gaussian probability density is
\begin{equation}
\label{eq:probdensity}
|\Psi_G|^2= \exp{\left[-\frac{d^3\vec{q}}{(2 \pi)^3}\left(A_{qR}\;\zeta_{\vq}\zeta_{-\vq} + B_{qR}\;\chi_{\vq}\chi_{-\vq} + C_{qR}\; \{ \zeta_{\vq}\chi_{-\vq}+\zeta_{-\vq}\chi_{\vq}\}\right)\right]}.
\end{equation}
Performing the functional integrals over the complex fields we get: 
\begin{eqnarray}
\label{eq:bispec}\nonumber
&&\langle \Psi | \zeta_{\vk_1}\zeta_{\vk_2} \zeta_{\vk_3} |\Psi\rangle =
12 (2\pi)^3 \delta\left(\sum_i \vk_i\right)\frac{1}{A_{1R}B_{1R}-C_{1R}^2}\frac{1}{A_{2R}B_{2R}-C_{2R}^2} \frac{1}{A_{3R}B_{3R}-C_{3R}^2}\\\nonumber
&& \Bigg[Z_R B_{1R}B_{2R}B_{3R} - 8\; Y_R C_{1R}C_{2R} C_{3R}+ 4 \sum_{l=1}^3 \sum_{i\neq l} \sum_{j\neq i,l} X_{lR} C_{iR}C_{jR}B_{lR} \\
&-&2 \sum_{l=1}^3 \sum_{i\neq l} \sum_{j\neq i,l} W_{lR} B_{iR}B_{jR}C_{lR} \Bigg]
\end{eqnarray}
where we only keep the lowest order term in the expansion parameter $\mu$ and $R$ denotes the real part of the various kernels. Note that $\langle \zeta_{\vk_1}\zeta_{\vk_2} \zeta_{\vk_3} \rangle$ has contributions from the spectator field $\chi^3$ term ($\propto Y$) and the mixed terms $\chi \zeta^2$ ($\propto W_i$) and $\zeta \chi^2$ ($\propto X_i$) thanks to the powers of the entanglement kernel $C_k$ multiplying each of them.  If the entanglement vanishes, i.e. $C_k=0$, we recover the standard BD bi-spectrum.

\section{\label{sec:3}
The Bi-Spectrum in the small entanglement strength limit}

Solving the equations for the kernels in the cubic wavefunction is generally a difficult undertaking. On the other hand, the fact that the BD state does give an extraordinarily accurate description of the CMB power spectrum tells us that the deviations from the BD state should be small. We make use of this fact to simplify our calculations by perturbing in the entanglement parameter $\lambda_k$ and keeping only the lowest non-trivial order result for the expectation value in eq.\eqref{eq:bispec}.

We start by expanding the mode functions found from the quadratic wavefunctional in terms of $\lambda_k^2$ as dictated by the right hand side of the mode equations in eq.\eqref{eq:riccatticonvert}, since $C_k\propto \lambda_k$. We thus write

\begin{equation} \label{exp1}
f_k =f_k^{BD}(\tau) \left(1+\lambda^2 \mathcal{F}_k(\tau)\right), \quad g_k =g_k^{BD}(\tau) \left(1+\lambda^2 \mathcal{G}_k(\tau)\right)
\end{equation}
%where  $\mathcal{F}_k(\tau)$ and $\mathcal{G}_k(\tau)$ are time and $k$ dependant functions whose solutions can be found analytically for the massless scalar case as shall be shown. 

The equations for the cubic kernels  can be simplified by making the following change of variables: 
\begin{eqnarray} \label{red1}
Z &=& \frac{\Zb}{\ft_1\ft_2\ft_3},\;\;\;\;\; Y = \frac{\Yb}{\gt_1\gt_2\gt_3}\\\label{red2}
W_{i} &=& \frac{\Wb}{\gt_i\ft_j\ft_l},\;\;\;\;\; X_{i} = \frac{\Xb}{\ft_i\gt_j\gt_l},  \;\;\;\;\;i,j,l=1,2,3,
\end{eqnarray}
where $\ft_k, f_k$ and $\gt_k,g_k$ mode functions are related to each other by $f_k =\alpha \ft_k$ where $\alpha = \sqrt{2 \epsilon} a $  and $g_k =  a\gt_k$ .

%\footnote{The BD curvature perturbation mode function will be: 
%\begin{equation}
%f_{BD} = -i  \frac{ (1- i k \tau) }{ \sqrt{2 k^3} \tau} e^{i k \tau};  \quad \tilde{f}_{BD} =  -\frac{ i H}{ \sqrt{4 \epsilon k^3}} (1- i k \tau) e^{i k \tau}    \rightarrow f_{BD} = \alpha \tilde{f}_{BD} 
%\end{equation} 
%while for the spectator filed. Note that between the two: $f_{BD} = g_{BD}$ and $\tilde{g}_{BD} =  \sqrt{2 \epsilon} \tilde{f}_{BD}$. :
%\begin{equation}
%g_{BD} = -i  \frac{ (1- i k \tau) }{ \sqrt{2 k^3} \tau} e^{i k \tau}  ;  \quad \tilde{g}_{BD} = - \frac{ i H}{ \sqrt{2 k^3}} (1- i k \tau) e^{i k \tau}    \rightarrow g_{BD} = a \tilde{g}_{BD}. 
%end{equation}}.
Using eqs.\eqref{red1}-\eqref{red2} the equations of motion for the cubic kernels simplify to:
\begin{eqnarray}\label{ode_bar1}
i \bar{Z}^{\prime} &=& \lambda \sum_i \frac{\bar{W}_i }{g^2_i} + \tilde{f}_1 \tilde{f}_2 \tilde{f}_3 \left[ \mathcal{S}_Z^{(0)} + \mathcal{S}_Z^{(2)} \right]\\ \label{ode_bar2}
i \bar{X}_i^{\prime} &=& - \lambda  \left( \frac{\bar{Y} }{g^2_i}+\sum_{i \neq j\neq l} \frac{\bar{W}_j}{f_l^2} \right) +  \tilde{f}_i \tilde{g}_j \tilde{g}_l\; \left[\mathcal{S}_{X_i}^{(0)}+ \mathcal{S}_{X_i}^{(2)}\right]
\\\label{ode_bar3}i \bar{Y}^{\prime} &=& \lambda \sum_i \frac{\bar{X}_i }{f^2_i} + \tilde{g}_1 \tilde{g}_2 \tilde{g}_3\; \mathcal{S}_Y^{(1)}\\\label{ode_bar4}
i \bar{W}_i^{\prime} &=& - \lambda  \left( \frac{\bar{Z} }{f^2_i}+\sum_{i \neq j\neq l} \frac{\bar{X}_j}{g_l^2} \right) +  \tilde{g}_i \tilde{f}_j \tilde{f}_l\; \mathcal{S}_{W_i}^{(1)}.
\end{eqnarray}
The superscripts in the source terms indicate the power of $\lambda$ associated with that source term (see Appendix \ref{Apx1} for the explicit form). When there is no entanglement, i.e. $\lambda =  0$ the only two kernels that have a source term generated by their respective cubic Hamiltonians at zeroth order in $\lambda_k$ are $\Zb$ and $\Xb_i$, while $\Yb$ and $\Wb_i$ will vanish at this order\footnote{Note, however, that while the source term for $X_i$ ($\propto \zeta \chi^2$) does not vanish when there is no entanglement, it will no longer contribute to the $\langle\zeta^3\rangle$ bi-spectrum since it is multiplied by the entanglement parameter $C_k$ as can be seen in eq.\eqref{eq:bispec}. In this case it will only contribute to the $\langle \zeta \chi^2 \rangle$ bi-spectrum.}:
 \begin{eqnarray}
i \bar{Z}_{BD}^{\prime} &=&  \tilde{f}_1 \tilde{f}_2 \tilde{f}_3 \left[ \mathcal{S}_Z^{(0)}  \right]\\ 
i \bar{X}_{i\;BD}^{\prime} &=& \sum_{i \neq j\neq l} \tilde{f}_i \tilde{g}_j \tilde{g}_l\; \left[\mathcal{S}_{X_i}^{(0)}\right]\\
i \bar{Y}_{BD}^{\prime} &=& 0 \\
i \bar{W}_{i\;BD}^{\prime} &=& 0
\end{eqnarray}
We can then expand $\bar{Z},\bar{Y},\bar{W}_i,\bar{X}_i$ in powers of $\lambda$:
\begin{eqnarray}
\Zb(\tau) &=&\Zb^{BD}(\tau) +\lambda\Zb^{(1)}(\tau) +\lambda^2 \Zb^{(2)}(\tau) +\mathcal{O}(\lambda^3),\\
\Xb_i(\tau) &=& \Xb_i^{BD}+\lambda \Xb_i^{(1)}(\tau) +\lambda^2 \Xb_i^{(2)}(\tau) +\mathcal{O}(\lambda^3)\\
\Yb(\tau) &=&\lambda \Yb^{(1)}(\tau) +\lambda^2 \Yb^{(2)}(\tau) +\mathcal{O}(\lambda^3)\\
\Wb_i(\tau) &=& \lambda \Wb_i^{(1)}(\tau) +\lambda^2 \Wb_i^{(2)}(\tau) +\mathcal{O}(\lambda^3).
\end{eqnarray}
where, without loss of generality, we have set $\Yb^{BD}=0$ and $\Wb_i^{BD}=0$ to satisfy the relations above. 

Using these expansions in eqns.\eqref{ode_bar1}-\eqref{ode_bar4} and matching powers of $\lambda$ yields a series of equations for the entanglement part of the cubic coefficients ($\Zb^{(1)}, \Zb^{(2)}, \Xb_i^{(1)}...$). Each order $\mathcal{O}(\lambda^n)$ can be found by integrating over the solutions of the previous order:

\begin{eqnarray}
\mathcal{O}(\lambda^0)&:& i \bar{Z}_{BD}^{\prime} =\tilde{f}^{BD}_1 \tilde{f}^{BD}_2 \tilde{f}^{BD}_3 \mathcal{S}^{(0)}_Z, \quad i \bar{X}_{BDi}^{\prime} =\tilde{f}^{BD}_i \tilde{g}^{BD}_j \tilde{g}^{BD}_l \mathcal{S}^{(0)}_{X_i},    \\
\mathcal{O}(\lambda^1)&:& i \bar{W}_i^{(1) \prime} =   - \lambda\left( \frac{\bar{Z}_{BD}}{f^2_{BDi}} +  \sum_{i \neq j\neq l} \frac{\bar{X}^{BD}_j }{g^2_{BDl}} \right)+ \tilde{g}_i^{BD} \tilde{f}_j^{BD} \tilde{f}_l^{BD}   \mathcal{S}^{(1)}_{W_i},\\
\mathcal{O}(\lambda^2)&:& i \bar{Z}^{(2) \prime} =   \lambda^2 \sum_i \frac{\bar{W}^{(1)}_i }{g^2_{BDi}} + \tilde{f}^{BD}_1 \tilde{f}^{BD}_2 \tilde{f}^{BD}_3 \left[\lambda^2 \left(\sum_i F_i \right) \mathcal{S}^{(0)}_Z+ \mathcal{S}^{(2)}_Z\right].
\end{eqnarray}
Here we have only included the terms that will contribute up to $\mathcal{O}(\lambda^2)$ overall in the bi-spectrum.

Next, we expand the bi-spectrum in eq.\eqref{eq:bispec} in orders of the entanglement parameter $\lambda$. %This not only gives us the leading contribution of non-Gaussianity from entanglement, but can also help illustrate where the non-Gaussianity contributions come from at each order. 
The real part of the quadratic coefficients are\cite{Albrecht:2014aga}:
\begin{eqnarray}
A_{kR} &=& \frac{1}{2 |\ft_k|^2},\quad B_{kR} = \frac{1}{2 |\gt_k|^2},\\\;\;\nonumber\\
C_{kR}&=&\mathcal{R}e\left( \frac{\lambda}{\ft_k\gt_k}\right)=\lambda \frac{\cos(\theta_{kf}+\theta_{kg})}{|\ft_k||\gt|},
\end{eqnarray}
where each mode function $\ft_k = |\ft_k|e^{i\theta_{kf}}$ is expressed in terms of its magnitude and $k$-dependent phase $\theta_{kf}$.
Using these, the bi-spectrum of eq.\eqref{eq:bispec} can be rewritten:
\begin{eqnarray}
&&\langle \Psi | \zeta_{\vk_1}\zeta_{\vk_2} \zeta_{\vk_3} |\Psi\rangle = 12\ \delta\left(\sum_j \vk_j\right)  \prod_i \left(\frac{1}{1-4\lambda^2  \cos^2(\Theta_i)} \right)8 \Big[Z_R|\ft_1|^2|\ft_2|^2|\ft_3|^2 \\\nonumber
&&-2^2 \lambda \sum_l W_{lR} \cos(\Theta_l) |\ft_i|^2|\ft_j|^2 |\ft_l||\gt_l|+2^4 \lambda^2 \sum_l X_{lR}\cos(\Theta_l) \cos(\Theta_2) |\ft_i||\gt_i||\ft_j||\gt_j||\ft_l|^2 \\\nonumber
&&- 2^6 \lambda^3 Y_R \cos(\Theta_1)\cos(\Theta_2)\cos(\Theta_3)
|\ft_1||\gt_1||\ft_2||\gt_2||\ft_3||\gt_3|\Big]
\end{eqnarray}
where $\Theta_k\equiv \theta_{kf}+\theta_{kg}$ and again, $R$ denotes the real part of each coefficient. 
Using our expansion of the mode functions in powers of $\lambda_k$, we can deduce the form of the leading correction to the BD bi-spectrum:

%We now expand the mode functions around the BD bi-spectrum in orders of entanglement strength $\lambda$ to obtain the full perturbative form of the bi-spectrum up to second order in $\lambda$:
\begin{eqnarray}\label{bispec_lambda}
\langle  \zeta_{\vk_1}\zeta_{\vk_2} \zeta_{\vk_3} \rangle &=& 12 \delta\left( \sum_i \vec{k}_i\right) \Bigg[ Z_R^{BD}  |\tilde{f}^{BD}_1|^2  |\tilde{f}^{BD}_2 |^2  |\tilde{f}^{BD}_3|^2 \\\nonumber
&+& \lambda^2 \Bigg\{ Z_R^{BD} \left[\sum_i \left(2\mathcal{F}_{iR} + 4 \cos^2(\theta_i)\right) + Z_R^{(2)}\right]  |\tilde{f}^{BD}_1|^2  |\tilde{f}^{BD}_2|^2  |\tilde{f}^{BD}_3|^2 \\\nonumber
&-& 2^2 \sum_i \left[W^{(1)}_{i R}  \cos(\theta_i)  |\tilde{f}^{BD}_j|^2  |\tilde{f}^{BD}_l|^2  |\tilde{f}^{BD}_i|   |\tilde{g}^{BD}_i|  \right]  \\\nonumber
&+& 2^4  \sum_i \left[X^{(0)}_{i R}  \cos(\theta_j) \cos(\theta_l)  |\tilde{f}^{BD}_i|^2  |\tilde{f}^{BD}_j| |\tilde{g}^{BD}_j|  |\tilde{f}^{BD}_l|   |\tilde{g}^{BD}_l|  \right] \Bigg\}\Bigg]
\end{eqnarray}
At zeroth order the $Z^{BD}$ kernel reproduces the standard BD bi-spectrum. At quadratic order in $\lambda_k$, we see contributions coming from the higher order $Z^{(2)}$ term as well as the mixed field kernels $W_i$ and $X_i$. As previously mentioned, each scalar field contributes a copy of the $W_i$ and $X_i$ terms so that in the limit of large numbers of scalars (which do not have to be that large as we will see below) they will dominate. We will therefore focus on these contributions for the remainder of the paper.

\section{\label{sec:4} The Shape of the Entangled Bi-Spectrum}

%\subsection{Bi-spectrum: Amplitude and Shape}

We are now in a position to compute the $\zeta$ 3-point function taken in the entangled state, turn this into the bi-spectrum, which in its turn can be compared to the data obtained by the Planck satellite\cite{Ade:2015ava}.

%In order to systematically compare bi-spectra and constrain them with CMB data we can focus on three features: shape, amplitude and running.
%The shape describes the momentum dependence of the bi-spectrum, and therefore incorporates all the useful information that can help us distinguish between inflationary models. Each shape will have an amplitude that can be compared to data, and the running encodes the dependence of the bi-spectrum on the total momentum.

%The homogeneity and isotropy of the FRW background simplifies the momentum dependence of the bi-spectrum considerably. Because of homogeneity, the bi-spectrum forms a transitionally invariant triangle composed of the three momentum vectors $\bm{k}_1, \bm{k}_2,\bm{k}_3$. Mathematically this is expressed  by a delta function of the sum of the momenta. On the other hand, thanks to isotropy the triangle of momenta is also rotationally invariant and we can express the bi-spectrum as a function of only the magnitudes of the momenta. Given these simplifications 

Using the homogeneity and isotropy of the FRW background, the bi-spectrum can be expressed as follows,       
\begin{equation}
B_{\zeta}(\bm{k}_1,\bm{k}_2,\bm{k}_3) = (2\pi)^3 \delta(\bm{k}_1+\bm{k}_2+\bm{k}_3) \frac{\Delta_{\zeta}^2(k_*)}{(k_1 k_2 k_3)^2} \; S(k_1, k_2, k_3)
\end{equation}
where $\Delta_{\zeta}^2(k_*)$ is the dimensionless power spectrum evaluated at the end of inflation, and $S(k_1, k_2, k_3)$ is the shape function of the bi-spectrum. 
One final simplification can be used if the power spectrum is scale invariant (or nearly scale invariant). In this case, the shape function only depends on ratios of the magnitudes of the momenta:
\begin{equation}
x_2 = \frac{k_2}{k_1}, \quad x_3 = \frac{k_3}{k_1}.
\end{equation}
%The majority of the most common inflationary models satisfy these conditions, and therefore many of the standard tools  used to analyze non-Gaussianity, rely on them. 

In practice, when comparing to data, the bi-spectrum of a particular inflationary model has to be matched to a shape template. From a computational perspective, the easiest shapes to match to are those that can be factorized into separate functions for each of the momenta. This requirement can be relaxed at the cost of making the comparison more computational expensive.

The most standard templates are the equilateral ($k_1=k_2=k_3$), local, which is maximized in the squeezed limit ($k_3<<k_2\approx k_1$), and flattened ($k_1 \approx k_2+k_3$) triangle shapes. Other more complex templates for oscillatory and feature models have also been used in analyzing the Planck data \cite{Ade:2015ava}.

%For computational purposes these templates must be factorizable in the three momenta, however even when a theoretical shape template is not factorisable it may be possible to construct a factorisable template that approximates the theoretical one. 

%To determine how much the theoretical shape of a model overlaps with a particular shape template, one can implement several tools \cite{Babich:2004gb, } including a scalar product and a cosine of the distributions. 
 %The larger the overlap between the template and the theoretical shape, the more effectual the comparison with CMB data can be.  On the other hand, if the shape of a new model is considerably different then the standard templates used in the analysis, it cannot be compared to the data effectively. 
%The most standard templates are the equilateral ($k_1=k_2=k_3$), local, which is maximized in the squeezed limit ($k_3<<k_2\approx k_1$), and flattened ($k_1 \approx k_2+k_3$) triangle shapes. Other more complex templates for oscillatory and feature models have also been used in analyzing the Planck data \cite{Ade:2015ava}. 

As a first step, it is useful to look at one or more of the three triangle limits (equilateral, squeezed and flattened), as a way to ascertain if the bi-spectrum in question exhibits enhancements in any of these. Different enhancements in one or more of these limits can help distinguish different models of inflation and indicate which templates might be most effective to use. 

In what follows we will start by plotting the entangled shape function 
%proportional to slow roll parameter $\epsilon$ 
and compare it to the no-entanglement scenario. Then we will look more closely at the equilateral, flattened and squeezed limits of the entangled bi-spectrum. Generally, non-BD models will produce enhancements in the flattened limit as well as the squeezed limit \cite{Chen:2006nt,
Chen:2010xka}, and multi-field inflation models are usually characterized by large enhancements in the squeezed limit. Looking at these limits will therefore give us a first glance on possible distinguishing features between these models.

%\subsection{Entangled Bi-spectrum Shape}

\begin{figure}% 
    \centering
   \includegraphics[width=10cm]{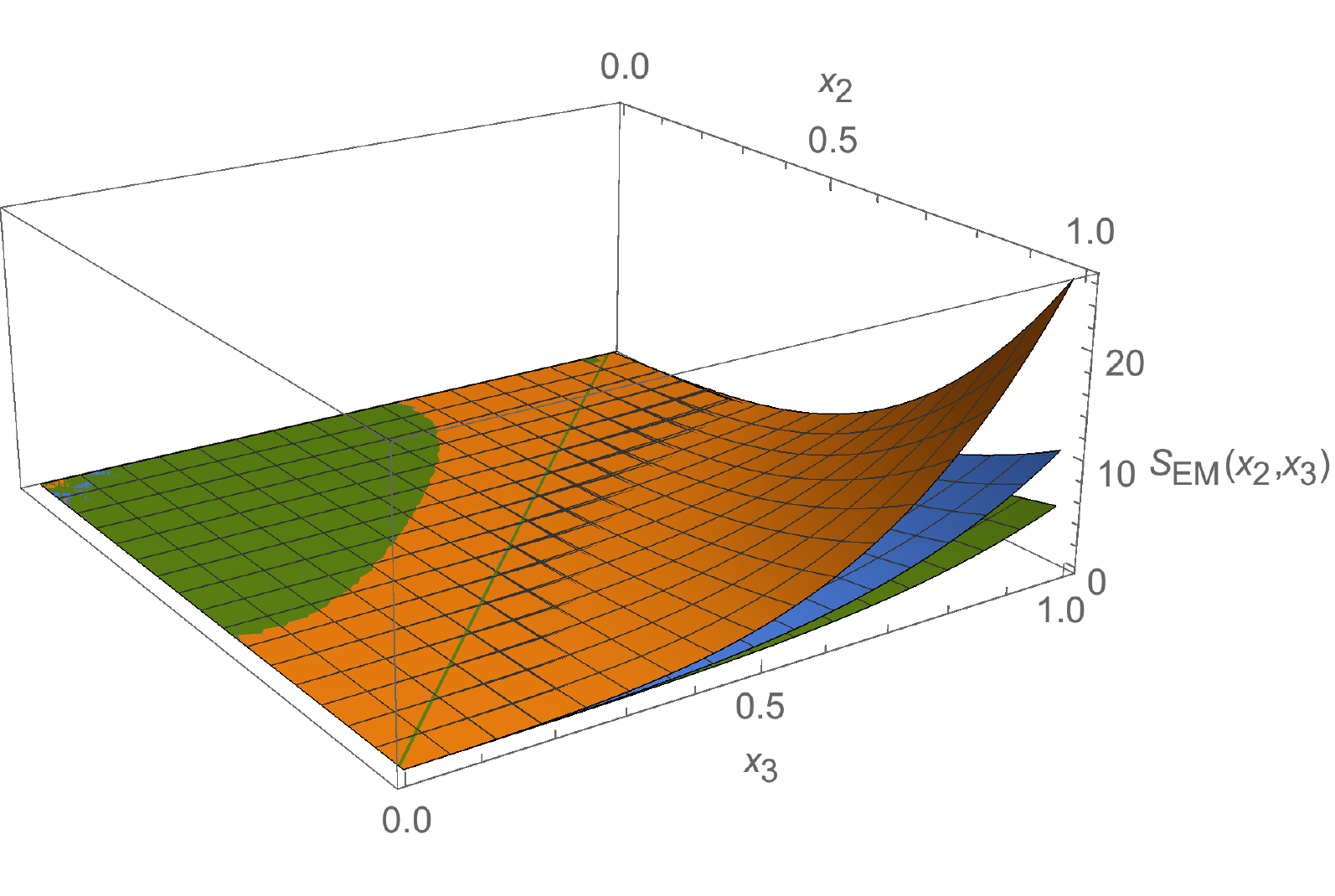}%
    \caption{The shape function  $x_2^2x_3^2 S_{\text{EM}}(x_2,x_3)$ of the entangled bi-spectrum vs. the no-entanglement BD bi-spectrum shape $x_2^2x_3^2 S_{Z}^{BD}(x_2,x_3)$. The lowest curve (green) is the BD shape, the middle curve (blue) is the entangled shape with one spectator field and the top curve (orange) is the entangled shape with four spectator fields. The initial time scale is set to $x_0=0.9$ which corresponds to a short, nearly minimal length of entangled inflation. The entanglement strength parameter is set to $\lambda=0.1$.}
    \label{shape_fig1}%
\end{figure}

\begin{figure}
    \centering
    \subfloat[$x_2^2x_3^2 S_{\text{EM}}(x_2,x_3)$ for $x_0=0.1$]{{\includegraphics[width=7cm]{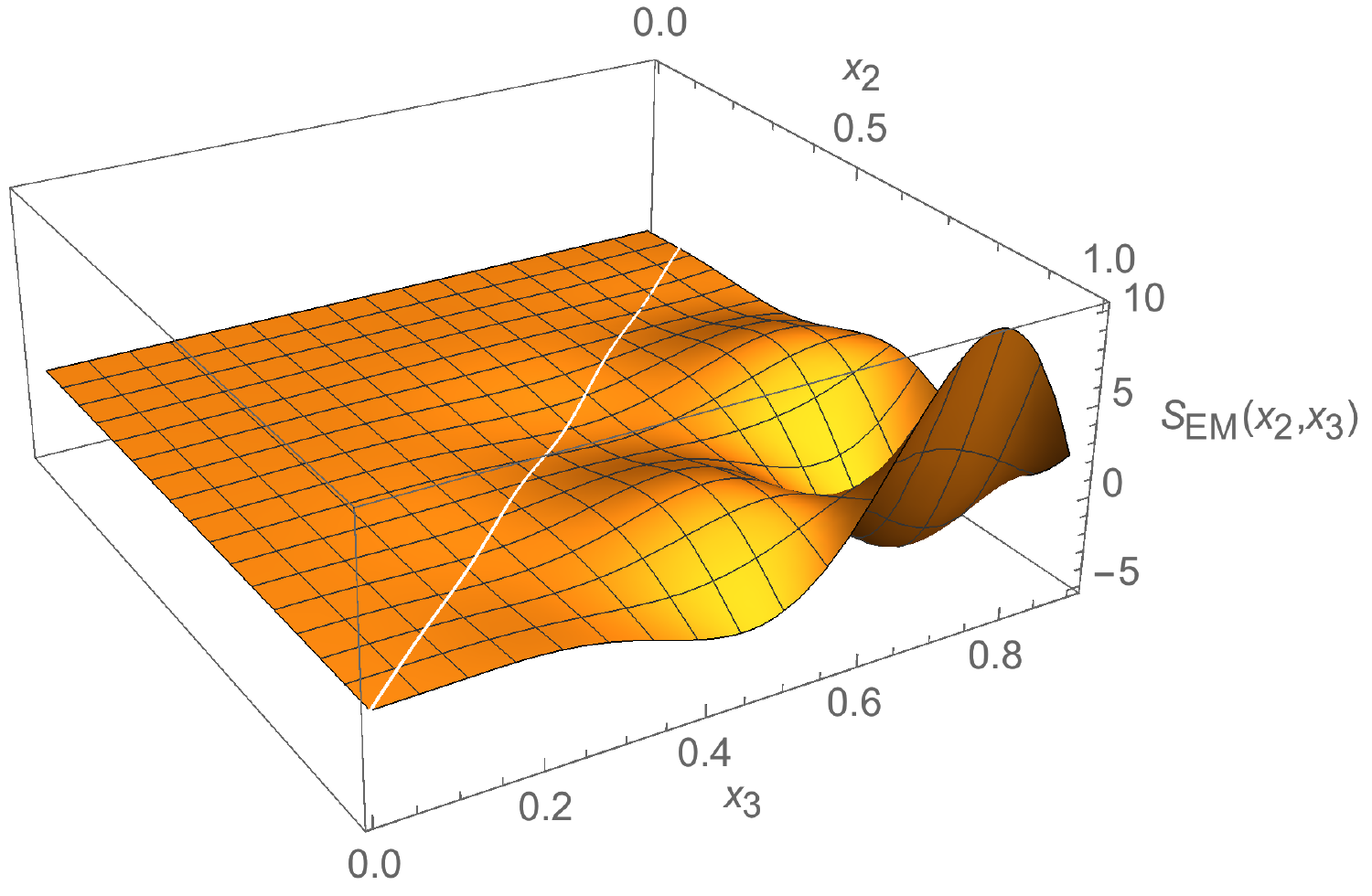}} }%
    \qquad
    \subfloat[$x_2^2x_3^2 S_{\text{EM}}(x_2,x_3)$ for $x_0=0.3$]{{\includegraphics[width=7cm]{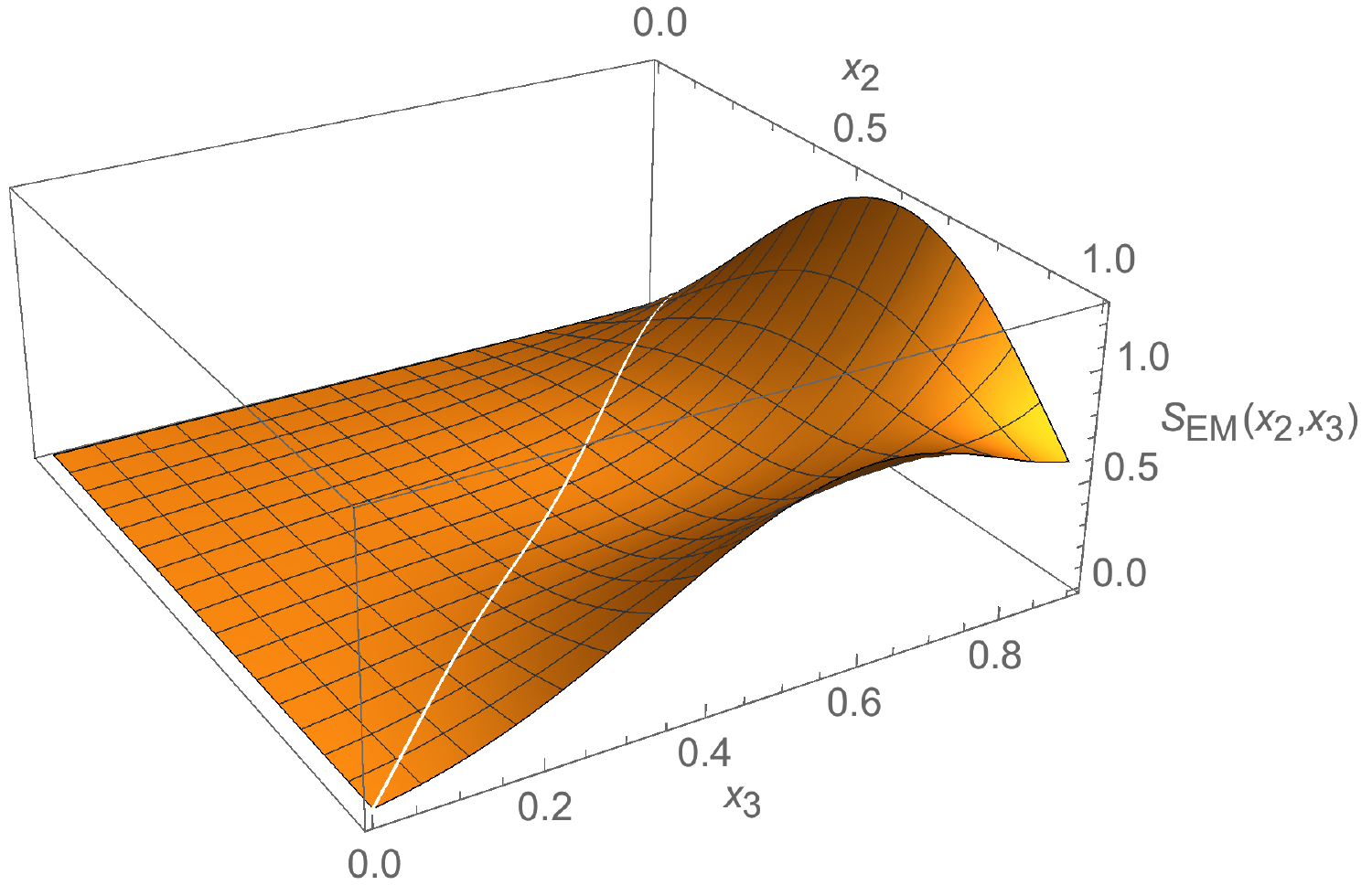} }}%
    \caption{The shape function $x_2^2x_3^2 S_{\text{EM}}(x_2,x_3)$ of the entangled bi-spectrum with one spectator field for long entangled inflation. In plot (a) $x_0=0.1$ and in plot (b) $x_0=0.3$. As the length of entangled inflation increases the shape functions starts exhibiting oscillatory features. The entanglement strength parameter is set to $\lambda=0.1$.}%
    \label{shape_fig2}
\end{figure}

%In this section we present the leading terms of the entangled bi-spectrum in comparison with the Bunch-Davies (no entanglement) bi-spectrum. In particular we focus on the mixed $\zeta \chi^2$ ($X_i$ kernel) and $\chi \zeta^2$ ($W_i$ kernel) terms which are not present in the standard scenario and do not affect the entangled two-point power spectrum. For multiple species of spectator fields these terms will dominate and can produce substantial enhancements in the bi-spectrum. 
In figures \ref{shape_fig1}-\ref{shape_fig2} we show plots of the shape function of the entangled bi-spectrum computed to $\mathcal{O}(\lambda^2$), and compare it to the no-entanglement BD shape. The shape function coming from the mixed $\zeta-\chi$ Hamiltonian is:
\begin{equation}\label{mixedshape}
    S_{\text{EM}}(x_2,x_3,x_0) = S^{BD}_Z(x_2,x_3) + n\;  \lambda^2 (S_W(x_2,x_3,x_0) + S_X(x_2,x_3,x_0))
\end{equation}
where $S^{BD}_Z$ is the shape function for the no-entanglement part, the subscript $EM$ denotes the entangled-mixed contribution to the bi-spectrum composed by the $S_W$ and $S_X$, the third and fourth line of the bi-spectrum in eq.\eqref{bispec_lambda}. Note that these terms are proportional to the number of spectator scalar fields $n$. In fig.\ref{shape_fig1} from bottom to top we have the no-entanglement BD shape (green), the entangled mixed shape for one scalar field (blue) and the same for 4 spectator fields (orange). For the purpose of comparison between these scenarios we have plotted all terms proportional to slow roll parameter $\epsilon$. Increasing the number of spectator fields can substantially increase the enhancements compared to the standard, no-entanglement scenario. These plots were done for a fixed value of $\lambda=0.1$. Since both the number of spectator fields $n$ and the entanglement parameter $\lambda^2$ multiply the new portion of the bi-spectrum, these two parameters are partially degenerate when varied over. However, since $\lambda<0.5$, enhancements that correspond to a larger combined value of $n \lambda^2$ can indicate the presence of multiple spectators in this scenario. 

The largest enhancement is in the equilateral limit. However the extra terms also contribute to the squeezed and flattened limits as will be shown in the next subsection. 
The finite time integral introduces a new scale $k_0$, which corresponds to the wavenumber that exits the horizon at the time entanglement turns on: $\tau_0$ ( i.e. $k_0=-1/\tau_0$). The corresponding dimensionless ratio is  $x_0\equiv k_0/k_1$. In what follows, $x_0=1$ corresponds to the minimal amount of entangled inflation, such that entanglement begins when the first visible modes exit the horizon. As $x_0$ decreases the period of entanglement increases, and finally the limit where $x_0\rightarrow 0$, corresponds to eternal inflation. We will see that, as expected, taking $x_0\rightarrow 0$ will introduce divergences in the bi-spectrum. In fig. \ref{shape_fig2} we show plots of the entangled mixed bi-spectrum eq.\eqref{mixedshape} for different initial entangling times: $x_0=0.3,0.1$. As the length of entangled inflation is extended, the shape of the bi-spectrum starts acquiring oscillatory features which could help put a bound on $x_0$. Note however, as discussed in \cite{Albrecht:2014aga} entangled inflation cannot be extended arbitrarily since it would cause divergences in the energy density. 

%\textcolor{orange}{Things to put in results section
%\begin{itemize}
%\item Plot of $Z_{BD}+ W_i+X_i$ contributions modulated by number of species. While entanglement goes as $\lambda^2$ and has to be $\lambda<0.5$, number of species increases enhancements linearly, so it can become more important even for smaller $\lambda$. This effect can only be seen with the bi-spectrum since it does not appear in the power spectrum.
%\item types of enhancements of these terms: flattened suppressed by time integral.The same holds for the  squeezed, where there are terms that would diverge but are regulated by the finite integral part. This is different then for multifield as far as i can tell, where squeezed limit is large, but perhaps i'm missing something. They usually calculate it with the $\delta N$ formalism but i think it should result in the same thing. So the interesting thing here is that entanglement in finite inflation, unlike multifield and how some people do non-BD does not have huge enhancements in these limits and therefore can't be excluded in the same way. 
%\item consistency relation:these terms add extra contributions that do not scale like $P_{k_1} P_{k_2} (1-n_s)$ so just from that i'd say its violated, but perturbatevly since $lambda^2$ is small. More species would violated more strongly.
%\end{itemize}}
\subsection{Equilateral, Flattened and Squeezed Limits}
In this subsection we look at the three triangle limits described above: equilateral, flattened and squeezed. %Enhancements in the equilateral limit are most common for higher derivative inflationary theories, the flattened triangle limit are characteristic for non-BD inflationary models \cite{Holman:2007na, Meerburg:2009ys, Chen:2006nt,Agullo:2011xv} and large enhancements in the squeezed limit appear in multifield models, among others.
The $W_i$, as well as $Z^{(2)}$, portion of the entangled bi-spectrum have terms proportional to:
\begin{equation}
\label{div}
\propto \frac{1}{(k_2+k_3-k_1)^p}
\end{equation} 
where $p$ is some power. At first glance it may look as though these terms will diverge in the flattened and squeezed limits, as $k_2+k_3\rightarrow k_1$ and $k_2\approx k_1, k_3\rightarrow 0$. In practice, such divergences arise when assuming a non-BD state back to the infinite past $\tau \rightarrow -\infty$; in other words, if we take the limit of the bi-spectrum integrals to be $\tau \in [-\infty, 0]$. However, realistically the entanglement behavior (or any non-BD state effect) cannot be pushed back to the infinite past, and therefore there should be a cutoff at large momenta where the entangled state is no longer valid and must be replaced.  One way of implementing this is by setting the bi-spectrum integration limits to $\tau \in [\tau_0, 0]$ starting from some finite time $\tau_0$ when entanglement begins. This finite integration exactly cancels out the divergences that would otherwise arise in the flattened and squeezed limits.   
An example of such a term in the $W_i$ kernel is:
\begin{equation}
       k_2^2\left( \frac{k_1 k_3}{(k_2+k_3-k_1)^2} + \frac{k_1 - k_3}{(k_2+k_3-k_1)}\right)  \left(1- e^{i (k_2+k_3-k_1) \tau_0}\right) 
\end{equation}
where this can be seen explicitly; the exponential factor from the finite time integration regulates this term, allowing the flattened and squeezed limits to be finite. This differs from other non-BD models \cite{Chen:2006nt} and some multifield models which have large enhancements in flattened and squeezed limits, and may help distinguish between these differing models.

\begin{figure}%
    \centering
    \subfloat[$S_{\text{EM}}(x_0)/S^{BD}_Z$ ratio for equilateral, flattened and squeezed limits.]{{\includegraphics[width=7cm]{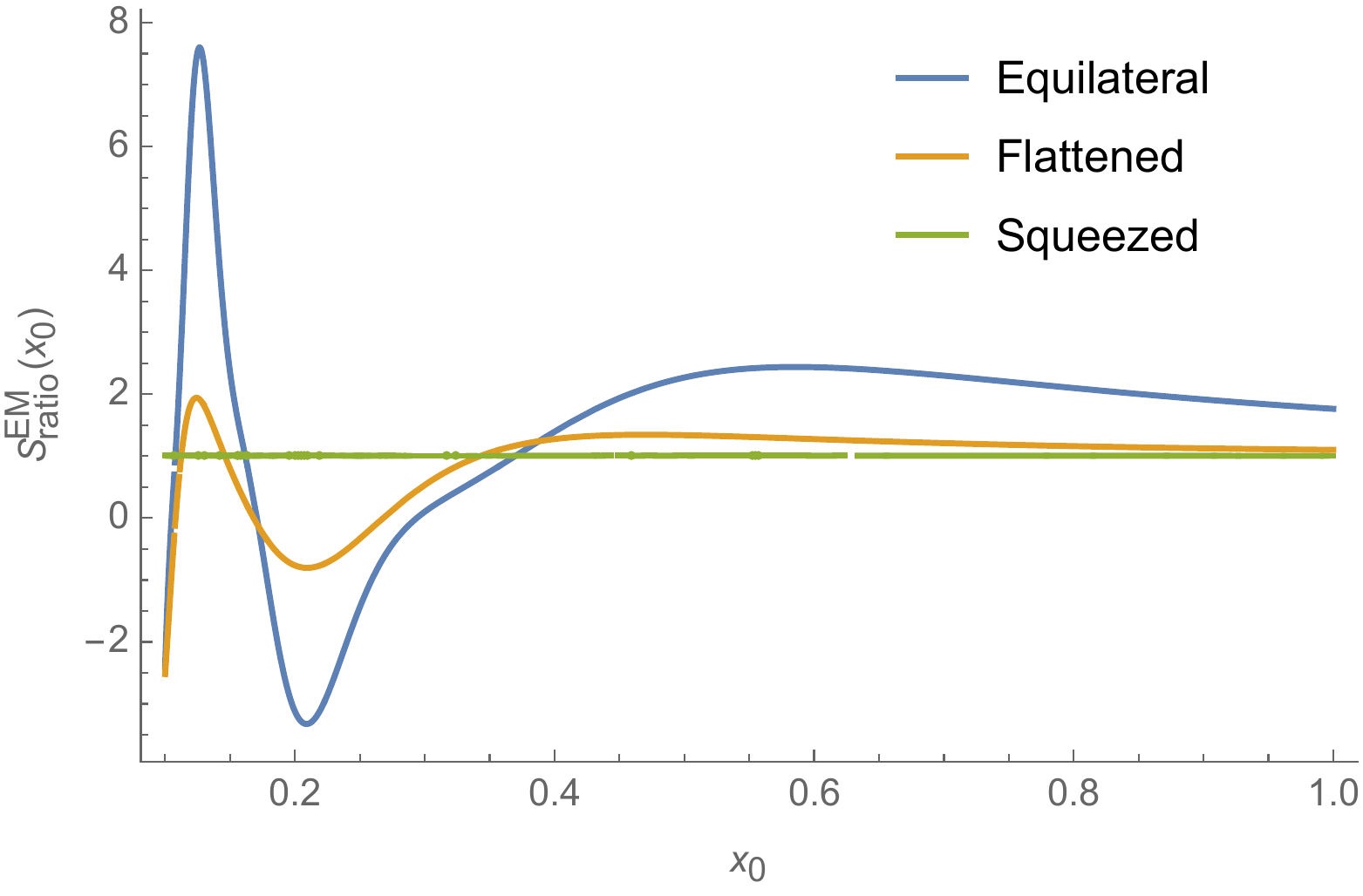}} }%
    \qquad
    \subfloat[$S_{\textrm{EM}}(x_0)/S^{BD}_Z$ ratio in the equilateral limit for multiple spectators. ]{{\includegraphics[width=7cm]{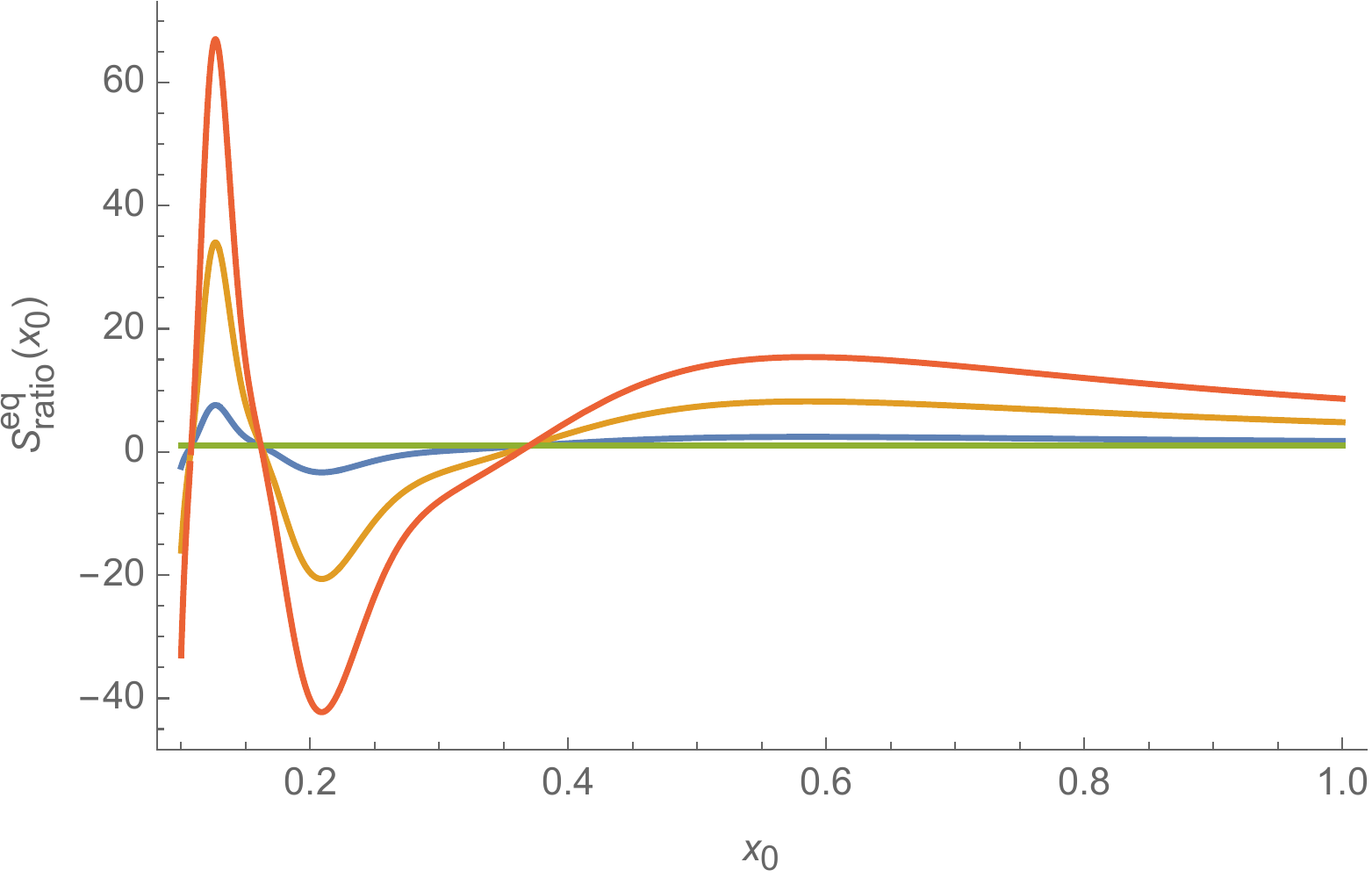}}}%
    \caption{(a) The ratio of the entangled mixed bi-spectrum over the no-entanglement BD bi-spectrum as a function of entanglement time parameter $x_0$ for the equilateral (blue), flattened (orange) and squeezed (green) limits. The enhancement with respect to the BD case is largest for the equilateral limit for most values of $x_0$. (b) The ratio of the entangled mixed bi-spectrum over the no-entanglement BD bi-spectrum as a function of entanglement time parameter $x_0$ in the equilateral limit for 1 (blue), 5 (orange) and 10 (red) spectator scalar fields. The green line shows the ratio of one with respect to the BD case.}%
    \label{shape_fig3}%
\end{figure}

\begin{figure}% 
    \centering
    \subfloat[$S_{\textrm{EM}}(x_0)/S^{BD}_Z$ ratio in the flattened limit for multiple spectators.]
    {{\includegraphics[width=7cm]{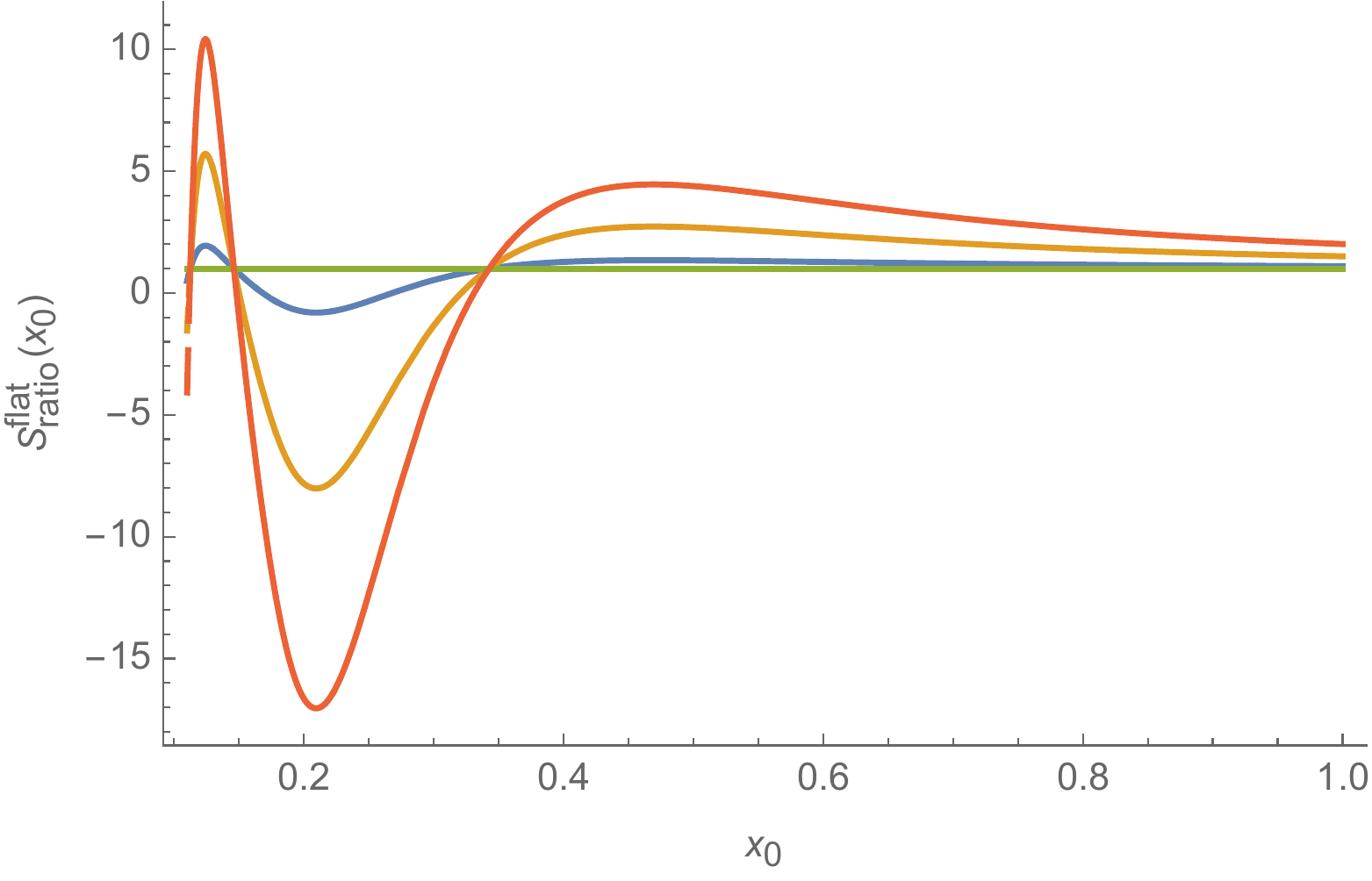}}}%
    \qquad
    \subfloat[$S_{\textrm{EM}}(x_0)/S^{BD}_Z$ ratio in the squeezed limit for multiple spectators.]{{\includegraphics[width=7cm]{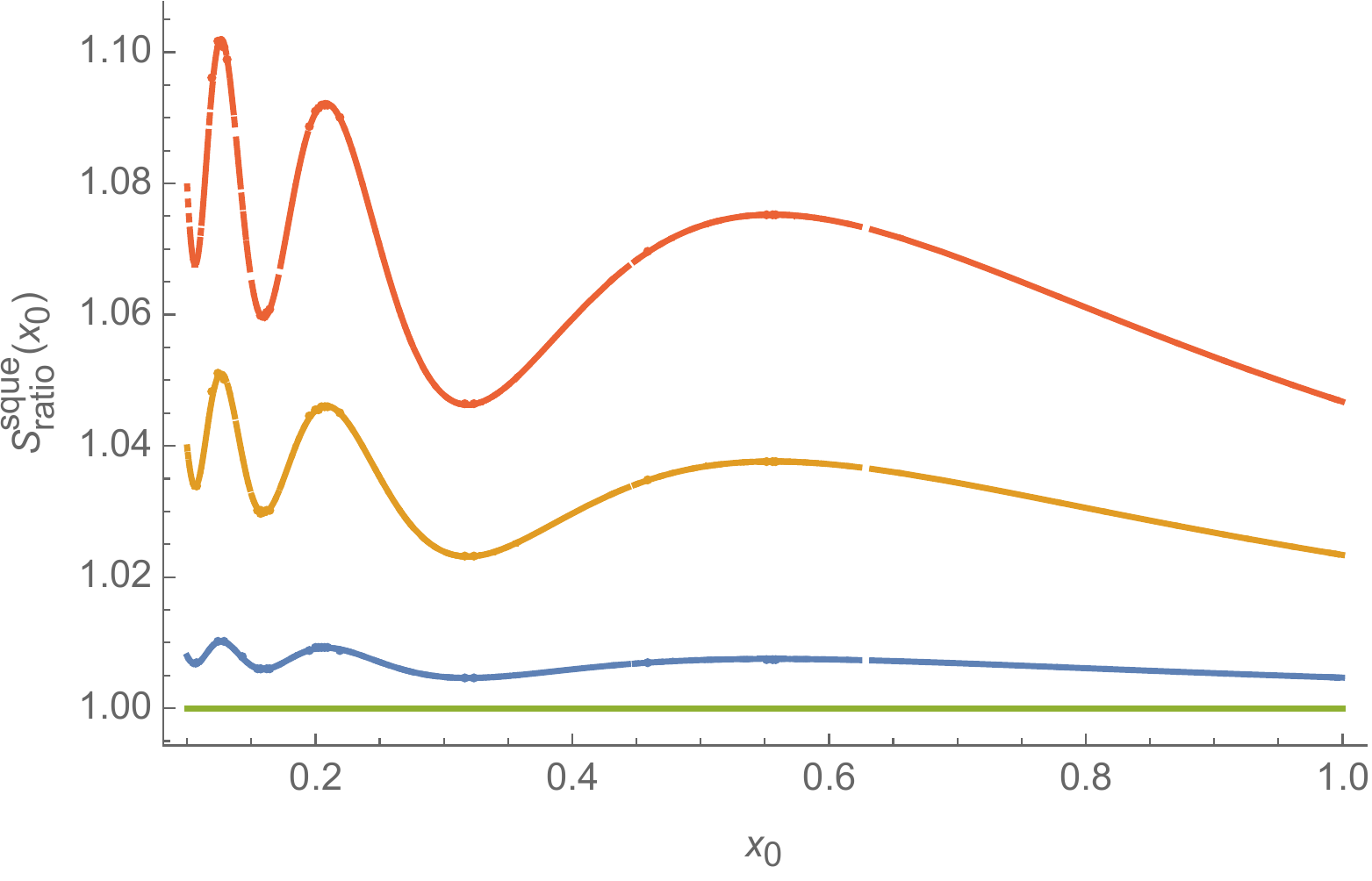}}}%
    \caption{The ratio of the entangled mixed bi-spectrum over the no-entanglement BD bi-spectrum as a function of entanglement time parameter $x_0$ in the flattened limit (figure (a)) and squeezed limit (figure (b)) for 1 (blue), 5 (orange) and 10 (red) spectator scalar fields. The green line shows the ratio of one with respect to the BD case.}%
    \label{shape_fig4}%
\end{figure}

While all the integrals are finite, the extra terms generated by entanglement will induce enhancements in both the flattened and squeezed limits. In fig.\ref{shape_fig4} we show examples of the flattened\footnote{In this plot we chose the flattened limit where $x_2=x_3\rightarrow 0.5$ as an illustrative example.} (a) and squeezed (b) limit of the ratio of the entangled bi-spectrum shape $S_{\text{EM}}$ and the  no-entanglement BD shape function $S_Z^{BD}$, for all initial entangling scales $x_0$.  The enhancements, compared to the standard scenario, are generally larger for the flattened limit rather then the squeezed limit for a given number of spectator fields. As mentioned above largest enhancements, however, are in the equilateral limit as can be seen in fig.\ref{shape_fig3}, with the three triangle limits compered directly in plot (a)  and with the equilateral limit for multiple spectator fields in plot (b).    

%Many non-BD models are tightly constrained by the current Planck non-Gaussinity data, with the largest signal resulting from a sinusoidal non-BD shape that has a $1.6\sigma$ for temperature $T$ raw significance, and $2.1\sigma$ for $T+E$ \footnote{Note that polarization constraints are considered to be preliminary at this stage.}. 

It is plausible that a full comparison of this entanglement model with current data could result in tight constraints on the entanglement strength parameter $\lambda$ and number of spectator fields $n$. We reserve a full analysis and comparison with data to future work. 
 
%\subsection{Squeezed limit and  Consistency Relation}
%In this section we examine the behavior of the entangled bi-spectrum in the squeezed triangle limit: $k_3<<k_1 \approx k_2$. One of the first shapes used to analyze non-Gaussianity was local non-Gaussianity. Local non-Gaussianity is produced by local interactions of a Gaussian distribution and is maximized in the squeezed limit. The local template is separable and takes the form:

%It was shown in \cite{Creminelli:2004yq} that single field inflation cannot produce large non-Gaussianity in the squeezed limit. In particular this powerful consistency relation \cite{Creminelli:2004yq} states that single field inflation, that starts in a BD vacuum, produces very small non-Gaussianity in the squeezed limit regardless of the exact form of the action. 
%On the other hand, multi-field inflation and non-BD inflation violate this consistency condition and can produce large non-Gaussianity signals in the squeezed limit [cite]. 

%We wish to ascertain whether the entanglement scenario violates the consistency relation as well, and if so, in what way. 
%On one hand, in this scenario, inflation is driven by only one field, since we treat the second field as a spectator. However, the entanglement forces these fields to not be in a BD vacuum so we might expect the consistency relation not to hold for this reason. Note that since we treat entanglement as a perturbative effect we may also reasonably expect the consistency relation to be broken only at a perturbative level. 

\section{\label{sec:5} Conclusion and Discussion}

In this work we calculated the bi-spectrum produced by entanglement between the curvature perturbations and a spectator scalar field during inflation. The bi-spectrum captures the higher order interactions of a theory and therefore can help further distinguish between inflationary models. 

Using the field theoretic Schr\"{o}dinger picture to describe the entangled state required the use of Schr\"{o}dinger perturbation theory to calculate the corrections to the $\zeta$ 3-point function. We found that terms in the cubic Hamiltonian that involve mixing between the spectator field $\chi$ and $\zeta$ were the most interesting ones to focus on, not least due to the fact that the effect they generate on the three point function can be enhanced by having many scalars entangle with the curvature perturbation. 

%The strength of the entanglement in this model is parametrized by a constant $\lambda$. Looking at the amplitude of the oscillations produced by the entanglement in the angular power spectrum, already constrains $\lambda$ to be small, in order to fit within the Planck data error bars. 
%We therefore calculated the entangled bi-spectrum perturbatively, expanding around the small parameter $\lambda$. 
%This allowed us to get an analytic solution to the cubic order equations of motion, and hence, an analytic expression for the shape of the entangled bi-spectrum. 

We argued that the entangled bi-spectrum shape, computed to first order in $\lambda$, differs from the non-BD model and the local shape non-Gaussianity typical of many multi-field models. In particular while these shapes differ moderately with respect to the short entangled inflation shape, more complex oscillatory shapes arise as entanglement is extended to the far past.

To further gain some intuition about the behavior of the entangled bi-spectrum we also looked at the equilateral, flattened and squeezed limits. In all three cases entanglement induces enhancements to the signal, with the largest being in the equilateral limit. The larger the entanglement parameter the larger the over all enhancement. Varying the length of entangled inflation will also affect the magnitude of these enhancements. For shorter entangled inflation ($x_0\gtrapprox 0.4$) the enhancements, for values of $\lambda$ that are not visibly excluded by angular power spectrum data ($\lambda \lessapprox 0.1$), get no larger then two times the BD bi-spectrum value in the case with one spectator field. However, for longer entangled inflation, and/or more spectator fields, the enhancements increase and could be used to exclude long entangled inflation and put a bound on the allowed number of entangled fields. In a full analysis to constrain this model, the effect of the tensor to scalar ratio $r$ (and other relevant cosmological parameters) will also have to be taken into account. 

The enhancement and features that appear in the entangled bi-spectrum shape can serve as a way to distinguish between this model and similar models, in a complementary way to the use of the angular power spectrum.

We conclude that the bi-spectrum provides an interesting set of new signals that expands the opportunities to test the initial state entanglement ideas beyond the signatures already explored in the power spectra~\cite{Albrecht:2014aga,Bolis:2016vas}.  These new signals allow us to further advance the goals of using inflation as ``the most powerful microscope in the Universe''~\cite{Albrecht:2018hoh} 
 to explore the nature of the initial state.

\acknowledgments

NB acknowledges funding from the European Research Council under the European Union's Seventh Framework Programme (FP7/2007-2013)/ERC Grant Agreement No. 617656 ``Theories and Models of the Dark Sector: Dark Matter, Dark Energy and Gravity.  AA thanks Department of Energy Grant DE-SC0019081 and UC Davis for support. RH thanks the Center for Quantum Mathematics and Physics and the Physics Department at UC Davis for hospitality while this work was in progress.

\appendix
\section{Source terms at each order in the entanglement \label{Apx1}}

Grouping together for each power of $\lambda$ the source terms generated by acting the $\zeta^3$ and $\zeta\chi^2$ cubic Hamiltonian on the Gaussian entangled state gives:
\begin{eqnarray}
\mathcal{S}^{(0)}_Z &=&  - \frac{\epsilon - \eta}{4 \epsilon a^2} \left\{A_2 A_3 +p.m.\right\} - a^2\frac{\epsilon(\epsilon +\eta)}{3} \frac{(k_1^2+k_2^2+k_3^2)}{2}\\\nonumber
&-& \frac{1}{a^2} \frac{1}{6} \left\{\frac{\vec{k}_2 \cdot \vec{k}_3 }{k_3^2} A_1 A_3 +p.m \right\} \\
\mathcal{S}^{(0)}_{X_i} &=& -  \frac{\epsilon }{ 2 a^2} \left\{B_j B_l \right\}-a^2\frac{\epsilon}{3}\frac{(k_1^2+k_2^2+k_3^2)}{2} - \frac{1}{2 a^2} \frac{1}{3} \left\{\frac{\vec{k}_i \cdot \vec{k}_l }{k_i^2} B_j A_i + \frac{\vec{k}_i \cdot \vec{k}_j }{k_i^2} B_l A_i \right\}\\
S^{(1)}_{W_i} &=& -\frac{\epsilon - \eta}{ \epsilon a^2} \left\{ A_j C_i + A_l C_i . \right\} - \frac{2}{a^2} \frac{1}{3} \left\{\frac{\vec{k}_l \cdot \vec{k}_j }{k_j^2} A_j C_i  +\frac{\vec{k}_j \cdot \vec{k}_l }{k_l^2} A_l C_i \right\} \\\nonumber
&-& \frac{\epsilon }{ 2 a^2} \left\{ B_i C_j + B_i C_l  \right\} - \frac{1}{2 a^2} \frac{1}{3} \left\{\frac{\vec{k}_l \cdot \vec{k}_j }{k_j^2}  A_j C_i + \frac{\vec{k}_j \cdot \vec{k}_l }{k_l^2} A_l C_i \right\}\\
\mathcal{S}^{(2)}_Z &=& - \frac{\epsilon}{2 a^2} \left\{  C_2 C_3 + p.m. \right\}
\end{eqnarray}
The zeroth order sources $\mathcal{S}^{(0)}_Z$ and $\mathcal{S}^{(0)}_X$ don't have any entanglement kernels $C_k$, the first order source $\mathcal{S}^{(1)}_W$ is proportional to one power of $C_k$ and the second order source $\mathcal{S}^{(2)}_Z$ has two powers of $C_k$.  

\bibliographystyle{JHEP}
%\bibliography{entang_bib}

\end{document}